\documentclass[pdflatex,sn-mathphys-num]{sn-jnl}

\usepackage{graphicx}%
\usepackage{multirow}%
\usepackage{amsmath,amssymb,amsfonts}%
\usepackage{amsthm}%
\usepackage{mathrsfs}%
\usepackage[title]{appendix}%
\usepackage{xcolor}%
\usepackage{textcomp}%
\usepackage{manyfoot}%
\usepackage{booktabs}%
\usepackage{algorithm}%
\usepackage{algorithmicx}%
\usepackage{algpseudocode}%
\usepackage{listings}%
\usepackage{lmodern}
\usepackage{anyfontsize} 
\usepackage{enumerate} 
\usepackage{booktabs}
\usepackage{rotating}
\usepackage{makecell} 
\begin{document}

\title{General Intellectual Humility Is Malleable Through AI-Mediated Reflective Dialogue}
\author{\fnm{Mohammad Ratul} \sur{Mahjabin}}\email{mohammadratul@usf.edu}
\author*{\fnm{Raiyan Abdul} \sur{Baten}}\email{rbaten@usf.edu}

\affil{\orgdiv{Computational Social Science and AI (CSSAI) Lab, Bellini College of Artificial Intelligence, Cybersecurity, and Computing}, \orgname{University of South Florida}, \orgaddress{\street{4202 E Fowler Avenue}, \city{Tampa}, \state{FL-33620}, \country{USA}}}

\abstract{General intellectual humility (GIH)—the recognition that one’s beliefs may be fallible and revisable—is associated with improved reasoning, learning, and social discourse, yet is widely regarded as a stable trait resistant to intervention. We test whether GIH can be elevated through a conversational intervention that combines staged cognitive scaffolding with personalized Socratic reflection. In a randomized controlled experiment ($N = 400$), participants engaged in a structured, LLM-mediated dialogue that progressed from conceptual understanding of intellectual humility to applying, analyzing, evaluating, and generating novel, self-relevant scenarios that instantiate it. Relative to a time-matched control, the intervention produced a systematic increase in GIH, reduced rank-order stability, and tripled the rate of reliable individual improvement. Crucially, these effects persisted over a two-week follow-up without detectable decay. The effects generalized across political affiliation and did not depend on baseline personality profile. These findings challenge the prevailing pessimism regarding the malleability of GIH and suggest that scaffolded, Socratic reflection delivered through structured dialogue can produce durable changes in general intellectual humility.
}

\keywords{General Intellectual Humility $|$ Large Language Models $|$ Behavioral Intervention}
\maketitle

\section*{Introduction}
General intellectual humility—the recognition that one’s beliefs may be fallible and revisable—has emerged as a promising antidote to a broad range of epistemic and societal challenges~\cite{porter2022clarifying,porter2022predictors,zmigrod2019psychological,leary2017cognitive}. Individuals higher in general intellectual humility (GIH) exhibit more accurate judgment and forecasting, greater openness to learning and discovery, higher-quality interpersonal and political dialogue, and lower susceptibility to polarization, extremism, and conspiracy beliefs~\cite{bowes2021looking,mellers2019forecasting,krumrei2025toward,porter2018intellectual,mahjabin2025wisdom}. These downstream outcomes are often studied separately across domains such as political behavior, scientific reasoning, education, leadership, and interpersonal relationships~\cite{smith2023you,bowes2022stepping,wong2021exploring,hoekstra2021aspiring,proma2025exploring}, yet they share a common upstream driver: the willingness to recognize the limits of one’s knowledge and revise beliefs accordingly. If so, cultivating GIH could represent a foundational intervention target capable of simultaneously mitigating multiple epistemic failures~\cite{scheffer2022belief}. This possibility raises a central question: can GIH itself be \textit{elevated}, potentially at scale?

In principle, there are reasons to believe that it could. GIH ultimately concerns beliefs about the fallibility of one’s own knowledge, and beliefs are known to change through learning, reflection, and experience~\cite{leary2018psychology}. Moreover, GIH correlates with broader psychological traits—such as openness to experience, epistemic curiosity, and need for cognitive engagement—that themselves exhibit some malleability over the life course and can respond to environmental and technological interventions~\cite{jackson2012can,stieger2021changing,leary2017cognitive,porter2018intellectual,roberts2008personality,roberts2005evaluating,stieger2018peach,fleeson2004moving}. Yet substantial evidence points in the opposite direction. Trait measures of GIH exhibit high test–retest reliability~\cite{porter2018intellectual,krumrei2016development}, and theoretical accounts suggest that it arises from relatively stable cognitive and motivational processes, potentially with evolutionary roots, making it difficult to change~\cite{church2016intellectual,leary2017cognitive,gregg2014intellectual,rieger2024potential, porter2020intellectual}. Indeed, while many beliefs change, beliefs about the reliability of one’s own beliefs—the \textit{metacognitive} core of intellectual humility—may be more resistant to revision. As a result, GIH is commonly treated as a durable epistemic disposition rather than a readily malleable psychological construct.

This pessimism has shaped the direction of intervention research. If raising GIH is difficult, a more feasible strategy is to target intellectual humility in specific downstream contexts or issues—specific intellectual humility (SIH)—which is more responsive to situational cues~\cite{hoyle2016holding,zachry2018situation,brienza2018wisdom,grossmann2016wise,priest2017intellectual,brienza2017social}. Correspondingly, much of the experimental literature focuses on \textit{momentary} or topic-specific manipulations, such as self-distanced reasoning about conflicts~\cite{grossmann2014exploring,kross2012boosting}, writing explanations that reveal gaps in one’s understanding of specific policies~\cite{fernbach2013political,johnson2016reflecting,meyers2020inducing}, reading about the benefits of IH~\cite{porter2020intellectual}, listing one’s “known unknowns” about a question~\cite{walters2017known}, and undergoing growth-mindset and self-affirmation tasks~\cite{porter2018intellectual,hanel2023using}. These approaches can temporarily nudge humility-related responses, but their effects are typically transient, context-dependent, or confined to specific issues rather than the broader disposition~\cite{smith2023you,hoyle2016holding,knochelmann2024effects}, with some findings showing limited replicability~\cite{crawford2021asking}. A few longer-duration interventions report encouraging results—such as self-distanced diary-keeping or guided conversations—but these effects appear contingent on specific situational conditions (e.g., relational affiliation must be established) and sustained, labor-intensive scaffolding~\cite{grossmann2021training,thorson2025increases}. Discouragingly, extended educational exposures have shown little to no success in shifting GIH~\cite{porter2022predictors}: for instance, a five-week philosophy course introduced students to IH without systematic success in elevating it~\cite{meagher2019intellectually}, and a week-long philosophy summer camp showed no measurable gain in students' IH~\cite{anderson2021development}. Consequently, the prevailing view is that while state-level, specific intellectual humility can fluctuate momentarily, general intellectual humility remains difficult to alter—let alone at scale.

Here, we question this conventional wisdom and ask whether the apparent immutability of GIH reflects the nature of the construct—or the limitations of how it has been targeted. If GIH is sustained by metacognitive processes governing how individuals evaluate knowledge, disagreement, and evidence, then shifting it may require interventions that systematically engage these processes through structured cognitive practice. Two ingredients are particularly relevant.

\textit{First}, a durable change in such metacognitive processes is unlikely to arise from isolated transient cues; rather, it can benefit from scaffolded engagement that progressively deepens how individuals understand and apply GIH in their own reasoning. Bloom’s taxonomy formalizes this progression by organizing learning into stages—from recalling and understanding a concept, to applying and analyzing it in concrete situations, and ultimately evaluating and generating new examples that integrate it into one’s thinking~\cite{bloom1956taxonomy,krathwohl2002revision,anderson2001taxonomy}. This progression may move learners beyond recognizing GIH as an abstract ideal to actively using it to interpret situations, weigh evidence, and guide belief revision \textit{across} contexts. 

\textit{Second}, Socratic questioning provides a mechanism for engaging these metacognitive processes by prompting individuals to examine assumptions, consider alternative viewpoints, and confront the limits of their knowledge~\cite{paul2007critical,vlastos1994socratic,king1994guiding}. When paired with later Bloom stages—where individuals analyze or generate scenarios—such questioning naturally personalizes the learning process: participants can draw on their own experiences, beliefs, and mental models to reason about when and how GIH applies. This personalized reflection may support generalization beyond static instruction, encouraging individuals to apply GIH within their own epistemic reasoning.

To operationalize these ingredients in a standardized, scalable, and interactive manner, we implemented a scaffolded conversational intervention using a large language model (LLM). LLMs enable adaptive, iterative conversation while preserving consistency across participants~\cite{costello2024durably,argyle2023leveraging}, allowing us to deliver Bloom-style scaffolding and Socratic probing at scale. In our intervention, participants engage in a structured dialogue that introduces three core facets of intellectual humility—awareness of the limits of one’s knowledge, openness to opposing viewpoints, and willingness to revise beliefs in light of evidence. The conversation progresses through Bloom-inspired stages that first reinforce conceptual understanding of these principles, then apply them to scenarios, and finally prompt participants to evaluate and generate diverse situations in which intellectual humility is beneficial. At each stage, the LLM provides brief guidance and poses targeted Socratic questions that build on the participant’s responses. This structure allows the conversation to remain standardized in design yet personalized in content, creating a brief but structured form of reflective engagement that helps us test whether scaffolded cognitive dialogue can elevate GIH in a measurable and durable manner.

We conducted a randomized controlled experiment in a virtual laboratory ($N = 400$). In the treatment condition, participants engaged in the Bloom-scaffolded Socratic dialogue focused on GIH. In the control condition, participants engaged in an unrelated but time-matched conversation. All participants completed a validated GIH questionnaire~\cite{leary2017cognitive} across three time points: pre-intervention (baseline), post-intervention, and a 14-day follow-up (see Methods and Materials for details). 

If GIH is meaningfully malleable under structured reflective engagement, three criteria should be met. First, the treatment should produce a GIH increase exceeding measurement error and retesting effects. Second, the effect should persist beyond the immediate post-intervention assessment. Third, changes should reflect structured cognitive alteration—evidenced by shifts in stability patterns and distributional properties—rather than transient response noise. We test these criteria directly.

\section*{Results}
\subsection*{In the Absence of Intervention, General Intellectual Humility Is Highly Stable}
The six-item GIH scale showed strong internal consistency at each time point (Cronbach’s $\alpha$ ranging from .86 to .90), indicating that the composite score reliably captured a coherent construct. We used this composite GIH score for our analyses. 

We first established a stringent stability benchmark. At the pre-intervention baseline, treatment and control groups did not differ in GIH ($\beta = -0.026$, $SE = 0.074$, 95\% CI $[-0.170, 0.118]$, $P = .73$), confirming successful randomization. Critically, in the control condition—where participants engaged in a time-matched but conceptually irrelevant conversation—pre–post test–retest reliability was extremely high ($r = .90$). This value indicates that participants’ relative standing in GIH remained nearly unchanged over time when no targeted intervention was implemented. In other words, absent intervention, GIH behaved like a stable trait. This high stability provides a demanding benchmark: any meaningful intervention must overcome strong rank-order persistence.

\subsection*{The Intervention Disrupts Stability and Produces a Systematic Causal Increase in GIH}
We next tested whether our conversational intervention altered this stability pattern. In the treatment group, pre–post test–retest reliability was lower ($r = .80$), indicating reduced preservation of rank ordering relative to the control group. This suggests that the intervention induced heterogeneous changes across individuals rather than a uniform shift in their GIH.

\begin{figure*}
    \centering
    \includegraphics[width=1\linewidth]{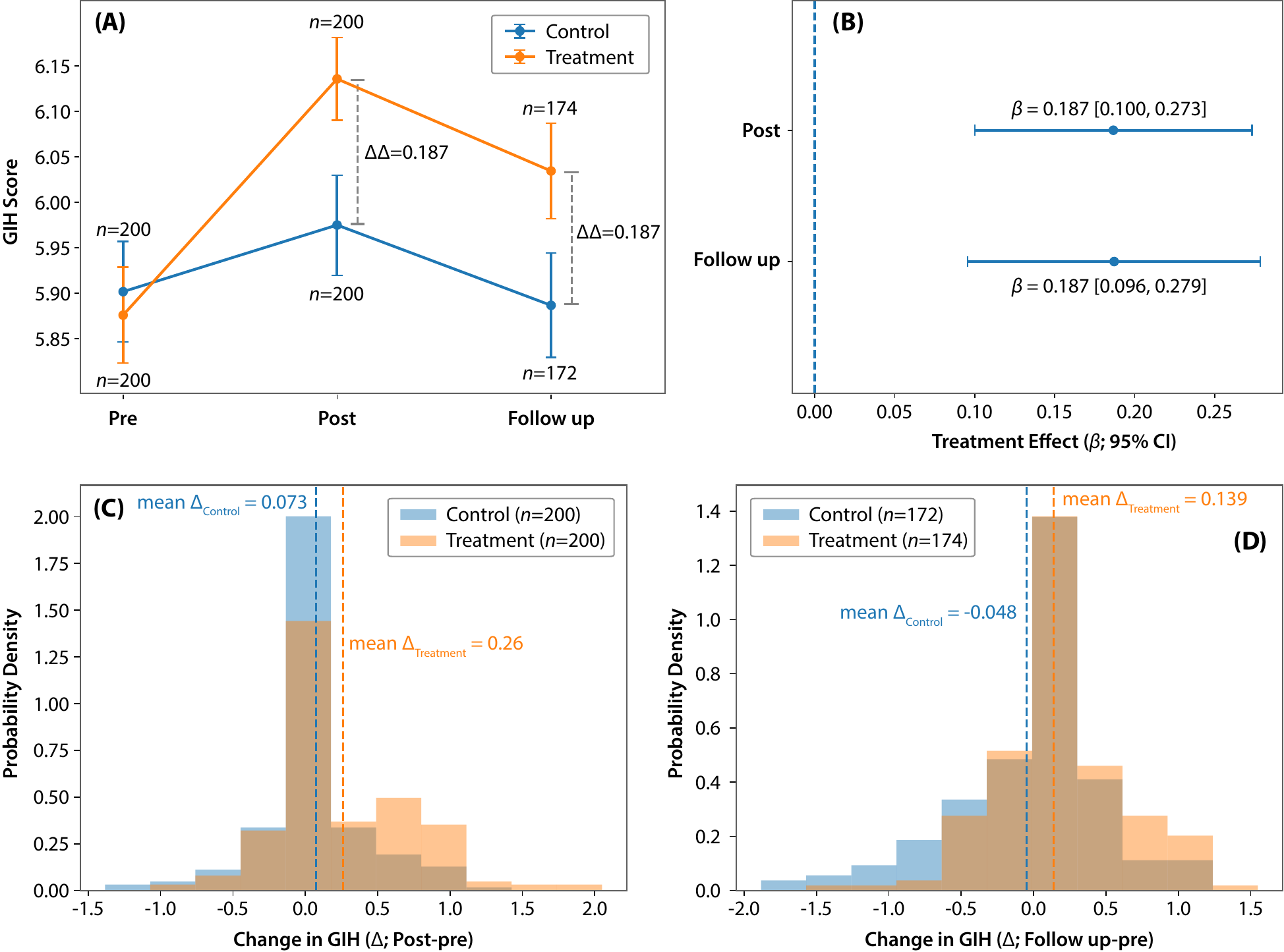}
    \caption{\textbf{(A) Mean GIH trajectories by condition.} Mean GIH across pre-intervention (baseline), immediate post-intervention, and 14-day follow-up, shown for treatment and control groups with standard errors of the mean. All available observations at each time point were used. Groups were comparable at baseline, diverged immediately after the intervention, and remained separated at follow-up. Dashed vertical segments mark the difference-in-differences (DiD) contrasts at post and follow-up. \textbf{(B) Mixed-model treatment effects.} Mixed-effects estimates of the treatment effect on GIH at immediate post-intervention and 14-day follow-up, shown as condition-by-time interaction coefficients. Both estimates are positive and statistically significant, and their near equivalence is consistent with persistence of the treatment effect without detectable decay. \textbf{(C)-(D) Distribution of individual changes in GIH by condition.} Left: immediate post-intervention change relative to baseline (post - pre). Right: 14-day follow-up change relative to baseline (follow-up - pre). Dashed vertical lines indicate group means. Across both cases, the treatment distribution is shifted rightward relative to the control distribution, indicating larger gains in GIH under the intervention. }
    \label{fig:composite}
\end{figure*}

To formally estimate mean-level change, we fitted a linear mixed-effects model with random intercepts for participants (800 observations; 400 participants). The interaction between condition and time was statistically significant ($\beta = 0.187$, $SE = 0.04$, $z = 4.67$, $P < .001$, 95\% CI $[0.108, 0.265]$, see Supplementary Table~\ref{tab:mixed_model_GIH_prepost}). This interaction term quantifies how much more the treatment group changed relative to the control. In change-score terms, the treatment group increased by $\Delta = 0.26$ points from pre to post, whereas the control group increased by $\Delta = 0.073$, yielding a difference-in-differences (DiD) estimate of $\Delta\Delta = 0.187$. This quantity corresponds to the model-based interaction estimate ($\beta = 0.187$), and indicates that the treatment group improved by approximately $0.19$ points more than the control group on a 7-point scale. In standardized terms, this corresponds to Hedges’ $g = 0.46$, a moderate effect size. Nonparametric tests yielded consistent evidence: a Mann–Whitney test on change scores indicated a significant between-group difference ($U = 24{,}744$, $P < 10^{-4}$), and a permutation test on the DiD (using observed $\Delta\Delta = 0.187$) likewise indicated a significant between-group difference ($P < 10^{-4}$). Taken together, the intervention both reduced rank-order stability and produced a moderate, statistically robust increase in mean GIH. These results provide initial evidence of the first criterion—demonstrating a statistically robust increase in GIH by disrupting its otherwise stable trait-like structure.

Mean differences alone cannot establish that individuals changed beyond measurement noise. To assess whether observed changes exceeded measurement noise at the individual level, we quantified Reliable Change Indices (RCI) using baseline reliability ($\alpha = 0.864$) and variability ($SD = 0.763$) (see Methods and Materials for details). The resulting standard error of the difference was $SE_{\text{diff}} = 0.398$, meaning that a change exceeding approximately 0.78 scale points (1.96 $\times SE_{\text{diff}}$) qualifies as statistically reliable improvement at the individual level. In the treatment group, $14.5\%$ of participants met this criterion, compared to $4.5\%$ in the control group. This corresponds to a risk ratio of $3.11$, indicating that participants in the treatment condition were more than three times as likely to show reliable improvement as those in the control condition. The corresponding odds ratio was $3.47$, indicating that the odds of meaningful improvement were more than tripled under treatment. Reliable decline was rare in both groups ($1.0\%$ in the treatment group and $2.5\%$ in the control). Thus, the intervention increased not only average GIH but also the probability of meaningful individual-level improvement beyond measurement error, further supporting the first criterion.

If an intervention simply added a constant to everyone’s score, variability would remain unchanged. Instead, we observed a significant increase in change-score variance under treatment: the standard deviation of post-intervention gain was $0.45$ in the treatment group versus $0.34$ in the control. A Brown–Forsythe test (a robust alternative to Levene’s test that is less sensitive to non-normality) comparing gain variances shows that the increase was significant ($F = 15.07$, $P < 10^{-3}$). This increase in variability indicates heterogeneous responsiveness: some individuals changed substantially, others minimally. Such structured heterogeneity is consistent with the elevated RCI responder rate and further supports that the intervention measurably altered individual trajectories. This pattern provides evidence for the third criterion—changes in distributional properties beyond uniform shifts—consistent with heterogeneous uptake of the cognitive processes engaged by the intervention.

\subsection*{The Treatment Effect Persists at 14 Days Without Detectable Decay}
We next evaluated durability using a three-timepoint mixed-effects model (1,146 observations; 400 participants). Significant condition $\times$ time interactions were observed both immediately post-intervention ($\beta = 0.187$, $SE = 0.044$, $z=4.22$, 95\% CI $[0.100, 0.273]$, $P < .001$) and at 14-day follow-up ($\beta = 0.187$, $SE = 0.047$, $z=4.01$, 95\% CI $[0.096, 0.279]$, $P < .001$; see Figure~\ref{fig:composite} and Supplementary Table~\ref{tab:mixed_model_GIH_followup}).

The follow-up effect was virtually identical to the immediate post-intervention effect. In change-score terms, this corresponds to a difference-in-differences (DiD) of $\Delta\Delta = 0.187$ at follow-up, matching the post-intervention estimate. The standardized effect size remained moderate (Hedges’ $g = 0.41$). A direct contrast test comparing the post and follow-up interaction coefficients revealed no detectable attenuation ($\beta_{\text{post}} - \beta_{\text{followup}} = -0.0005$, $SE = 0.0467$, $z=-0.011$, 95\% CI $[-0.0919, 0.0910]$, $P = .99$). The retention ratio (follow-up effect divided by post-effect) was $1.02$, indicating preservation of the treatment advantage relative to the control. Thus, the intervention-induced elevation in GIH persisted over two weeks without evidence of decay, satisfying the second criterion of durability. This persistence is consistent with engagement of cognitive processes beyond transient activation.

\subsection*{The Effect Generalizes Across Political Affiliation and Personality}
We examined whether treatment efficacy differed across self-reported political affiliations. Using heteroscedasticity-robust regression models predicting change scores (with baseline GIH included as a covariate), no condition $\times$ party interaction was statistically significant (all $P > .50$). For example, relative to Democrats (reference category), the treatment $\times$ Republican interaction was $\beta = 0.068$ ($SE = 0.1$, $z=0.65$, 95\% CI $[-0.136, 0.272]$, $P = .52$), and the treatment $\times$ Independent interaction was $\beta = -0.022$ ($SE=0.086$, $z=-0.258$, 95\% CI $[-0.191, 0.146]$, $P = .80$; see Supplementary Table~\ref{tab:party_moderation}). These findings indicate that the positive and significant treatment effect generalized across political groups, with no detectable moderation by partisan affiliation.

We next tested whether treatment efficacy varied across Big Five personality traits using heteroscedasticity-robust regression models predicting change scores (controlling for baseline GIH). No condition $\times$ trait interaction was statistically significant (all $P > .10$; see Supplementary Table~\ref{tab:big5_moderation}). For example, the treatment $\times$ Conscientiousness interaction—the largest in magnitude—was not statistically significant ($\beta = 0.057$, $P = .114$), and interactions for Openness ($\beta = -0.072$, $P = .217$), Agreeableness ($\beta = -0.003$, $P = .940$), Extraversion ($\beta = 0.031$, $P = .358$), and Neuroticism ($\beta = -0.047$, $P = .137$) were likewise not statistically significant. These results indicate that the intervention’s efficacy did not depend detectably on baseline personality, further supporting its generalizability across individual differences. This pattern is consistent with changes in a general epistemic disposition (GIH) rather than context-specific or trait-dependent effects.

As a validation check, baseline GIH showed associations with all Big Five traits in directions consistent with prior literature—most notably positive correlations with Openness (Spearman's $\rho = 0.28$, $P < .001$) and Agreeableness ($\rho = 0.21$, $P < .001$; see Supplementary Table~\ref{tab:big5_gih_correlations})~\cite{porter2022predictors}. Thus, while personality traits are themselves associated with baseline levels of GIH, they do not appear to moderate responsiveness to the intervention in \textit{changing} GIH.

The treatment effect remained positive and statistically significant in a covariate-adjusted model including baseline GIH, all five personality traits, conversational duration in seconds, and number of conversational turns ($\beta = 0.194$, $SE = 0.037$, $z = 5.22$, 95\% CI $[0.121, 0.267]$, $P < .001$; see Supplementary Table~\ref{tab:covariate_model_big5}). Conversational duration was not associated with change ($P = .66$), and the number of turns showed only a marginal association ($P = .074$). Among personality covariates, Conscientiousness was positively associated with change ($\beta = 0.078$, $P = .004$), whereas other traits were not statistically significant. Critically, the persistence of the treatment effect after accounting for baseline traits and engagement features indicates that the observed elevation in GIH cannot be attributed to differential engagement or pre-existing personality differences. Collectively, these analyses demonstrate that a brief scaffolded reflective conversation produces a moderate, reliable, durable, and broadly generalizable increase in GIH.

\section*{Discussion}
We report evidence that a brief, scaffolded reflective dialogue can produce a moderate, reliable, and durable increase in general intellectual humility (GIH)—a construct widely regarded as trait-like and difficult to alter. In the absence of intervention, GIH exhibited extremely high short-term stability, consistent with prior work treating it as a durable epistemic disposition~\cite{porter2018intellectual,krumrei2016development}. Yet under treatment, mean GIH increased, reliable individual improvement more than tripled relative to control, rank-order stability weakened, and change-score variance increased. Crucially, the treatment advantage persisted for at least two weeks without detectable decay. Taken together, these results challenge the prevailing pessimism in the literature. They suggest that prior null or weak findings may reflect not the immutability of GIH per se, but the limitations of intervention designs that rely on brief nudges, topic-bound prompts, or socially contingent scaffolds~\cite{porter2022predictors,hoyle2016holding,thorson2025increases}. More generally, our findings support the idea that when GIH is engaged as a metacognitive capacity—through scaffolded learning and personalized reflection—it can shift in measurable and durable ways.

This contribution is substantive for the broader literature in two respects. \textit{First,} at a \textit{scientific} level, the intervention combines two ingredients---Bloom-style cognitive scaffolding and Socratic questioning---that have not been jointly examined in prior research on intellectual humility or on LLM-mediated interventions. The Bloom progression structures engagement, moving participants from understanding GIH to evaluating and generating scenarios that instantiate it across contexts, while Socratic prompts personalize this trajectory by requiring individuals to interrogate their own assumptions, evidence, and reasoning. Together, these components target the metacognitive processes underlying GIH rather than treating it as a disposition that passively shifts under exposure or priming. This design also suggests a reverse pathway: whereas intellectual humility has been linked to epistemic curiosity and need for cognition—the motivation to think carefully, question assumptions, and pursue understanding~\cite{leary2018psychology,leary2017cognitive,porter2018intellectual}—repeatedly engaging individuals in such effortful, reflective reasoning may \textit{itself} cultivate GIH. Consistent with this possibility, our intervention repeatedly required participants to confront the limits of their knowledge and reason through competing perspectives—processes that are cognitively demanding and often attenuated under everyday constraints~\cite{scheffer2022belief}. In this sense, the present work contributes a framework—grounded in established learning and dialogic principles—for treating GIH as an educable metacognitive capacity. \textit{Second,} at a \textit{practical} level, prior successes have often depended on repeated diary-style practice or social conditions such as affiliation and trust~\cite{grossmann2021training,thorson2025increases}. Here, a relatively brief intervention produced durable gains without requiring sustained human facilitation or relational alignment, enabling scalable deployment. This creates a tractable pathway for embedding intellectual humility training in educational, organizational, and digital settings.

We assessed GIH using a validated self-report questionnaire rather than a behavioral task. Developing a valid behavioral measure of GIH, however, is conceptually and psychometrically challenging. Many candidate behaviors that appear to express general intellectual humility are multiply determined by correlated characteristics such as openness, agreeableness, submissiveness, conflict aversion, empathy, verbal fluency, or social strategy~\cite{leary2018psychology,porter2022predictors,mccrae2009openness,bowes2022stepping,spiegel2012open}. For example, reacting politely during disagreement may stem from intellectual humility, but it may also reflect a desire to avoid conflict; seeking information may reflect intellectual humility, but also curiosity, anxiety, or conscientiousness. Behavioral responses are further shaped by topic, stakes, audience, relationship, and situational norms. As a result, a single task can easily become a measure of social performance in a specific context rather than of general intellectual humility as a cross-situational epistemic disposition. Existing task-like measures are rare and often tailored to narrow populations or topic-specific expressions of intellectual humility~\cite{danovitch2019intelligence}. This creates a major opening for future work: if GIH is to mature as a scientific construct, the field will need a new generation of behavioral paradigms—or a calibrated task battery—that can isolate general epistemic fallibility recognition from its many correlates, demonstrate reliability across contexts, and establish convergent and discriminant validity.

In the absence of such behavioral measures, self-report questionnaires remain the primary tool for assessing GIH in the literature~\cite{meagher2019intellectually,anderson2021development,porter2022predictors}. Their usefulness depends on whether respondents can accurately report their epistemic tendencies and whether responses are unduly influenced by self-presentation concerns. In particular, items that directly ask individuals to rate their own intellectual humility may invite recall errors and socially desirable responding~\cite{abrahams2019social,duckworth2015measurement}---so, instead, validated questionnaires probe tendencies that reflect intellectual humility more indirectly. Evidence suggests that such measures exhibit acceptable reliability and are not unduly driven by social desirability biases~\cite{leary2018psychology,leary2017cognitive}. Moreover, scores on trait-level GIH measures correlate with behaviors central to intellectual humility, including information-seeking, cognitive flexibility, acknowledgment of intellectual limitations, and argument evaluation, implying external validity~\cite{leary2017cognitive,zmigrod2019psychological,porter2018intellectual,meagher2021intellectual}. The present findings should be understood as evidence that GIH—as captured by validated psychometric instruments—can be experimentally shifted, while highlighting the importance of developing complementary behavioral measures.

This measurement challenge is closely related to the nature of the construct itself. GIH is, at its core, a cognitive and metacognitive phenomenon at the individual level~\cite{leary2018psychology,church2017intellectual}. Social interaction may shape when intellectual humility is expressed, but the construct centrally concerns how individuals represent the limits of their knowledge, evaluate disagreement, and revise beliefs in light of evidence. This distinction is important for interpreting the present intervention. The LLM-based dialogue likely reduced barriers common in human interactions, such as identity threat, fear of embarrassment, or defensive responding in adversarial contexts~\cite{grossmann2014exploring,krumrei2020intellectual}. Prior work suggests that individuals may disclose uncertainty and engage challenging ideas more readily with AI than with human interlocutors~\cite{kim2025conversational}. But that does not make the present findings merely an artifact of a low-threat social setting. Rather, it highlights a useful property of the design: our approach isolates and trains the cognitive core of GIH without requiring the interpersonal trust and affiliation on which some social interventions depend~\cite{thorson2025increases}. A natural next step is therefore to examine whether gains in GIH transfer from such low-threat reflective settings to higher-stakes interpersonal and collective contexts.

Several features of the results increase confidence in the substantive importance of the effect. The treatment generalized across political affiliations and did not depend on baseline Big Five attributes, suggesting that the intervention was not narrowly tailored to a single ideological subgroup or personality profile. The effect also remained significant after adjusting for baseline GIH, personality traits, conversation duration, and number of turns, making it unlikely that the results were driven merely by time-on-task or superficial engagement intensity. At the same time, the increase in gain variance and the Reliable Change Index patterns indicate meaningful heterogeneity in responsiveness. Not everyone changed equally. This heterogeneity is theoretically informative: if GIH is sustained by metacognitive processes, then individuals may differ in readiness, fit with the intervention, prior epistemic habits, or susceptibility to reflective scaffolding. Future work should identify these moderators more precisely.

Several limitations should also be noted. The sample was drawn from a U.S.-based population, limiting claims about cross-cultural generalizability. Because humility norms, conversational styles, and epistemic values vary across societies, replication in non-U.S. and multilingual contexts is important. The follow-up window was limited to 14 days, so longer-term durability remains unknown. Finally, although the present evidence is consistent with structured cognitive engagement as the active ingredient, our design does not fully isolate which components of the intervention—Bloom scaffolding or Socratic probing—were necessary or sufficient. Component ablation studies would be especially valuable.

These findings point to a broader implication for the study of epistemic behavior. If general intellectual humility functions as an upstream driver of how individuals evaluate information, engage disagreement, and revise beliefs, then even modest shifts in GIH may influence downstream processes such as information-seeking, confidence revision, argument evaluation, and discourse quality across domains, including political reasoning, misinformation, and scientific judgment~\cite{bowes2021looking,mellers2019forecasting,smith2023you,hoekstra2021aspiring}. Although the present results do not directly establish these downstream effects, they make them empirically tractable. Indeed, our findings suggest that interventions targeting foundational metacognitive processes—rather than isolated beliefs or topics—may offer a scalable path for addressing multiple epistemic challenges through a common mechanism.

\section*{Methods and Materials}
The study protocol was approved by the Institutional Review Board at the University of South Florida. All participants provided informed consent prior to participation.

\subsection*{Participants and Procedure}
We recruited $N = 400$ participants via Prolific who were at least 18 years old, fluent in English, and residing in the United States. Participants were randomly assigned to either a treatment condition ($n = 200$) or a time-matched control condition ($n = 200$). Supplementary Table~\ref{tab:s_demographics} gives a detailed breakdown of participant demographics. Across demographic attributes (age, race, Hispanic origin, gender, education, occupation, and party affiliation), chi-square tests did not indicate any statistically significant differences between treatment and control conditions at the .05 level. Hence, any observable difference between the two conditions cannot be attributed to differences in the distributions of these demographic attributes.

The study followed a three-wave longitudinal design with assessments at (i) pre-intervention baseline, (ii) immediate post-intervention, and (iii) a 14-day follow-up. Complete data across all three time points were available for $N = 346$ participants ($n = 172$ control; $n = 174$ treatment). Attrition rates were comparable across conditions.

At baseline, participants completed measures of demographic characteristics, personality traits using the Big Five Inventory–2 short form (BFI-2s)~\cite{soto2017short,soto2017next}, and GIH~\cite{leary2017cognitive} (See Supplementary Text for full instruments). Participants then engaged in a text-based conversational task with a large language model (GPT-4.1), with content determined by experimental condition. In the treatment condition, participants engaged in a structured reflective dialogue designed to cultivate GIH through Bloom-scaffolded reasoning and Socratic questioning. In the control condition, participants engaged in an unrelated but time-matched conversation, controlling for interaction format and duration without targeting intellectual humility (see Supplementary Text for the full LLM system implementation details). GIH was reassessed immediately after the conversation and again at the 14-day follow-up.

\subsection*{Measures}
\textbf{General intellectual humility (GIH).} GIH was assessed using a validated six-item Likert-scale instrument (7-point scale)~\cite{leary2017cognitive}. Items capture beliefs and tendencies related to recognizing the limits of one’s knowledge, openness to opposing viewpoints, and willingness to revise beliefs. Composite GIH scores were computed as the mean of the six items, with higher values indicating greater intellectual humility.

\textbf{Reliable Change Index (RCI).} To assess whether individual-level changes exceeded measurement error, we computed the Reliable Change Index (RCI), defined as:
\[
RCI = \frac{X_{\text{post}} - X_{\text{pre}}}{SE_{\text{diff}}},
\]
where $X_{\text{pre}}$ and $X_{\text{post}}$ denote an individual’s pre- and post-intervention scores. The standard error of the difference is: $SE_{\text{diff}} = \sqrt{2} \times SEM$ and the standard error of measurement is: $SEM = SD_{\text{pre}} \sqrt{1 - \alpha}$, where $SD_{\text{pre}}$ is the baseline standard deviation and $\alpha$ is Cronbach’s alpha of the scale. Individuals with $|RCI| > 1.96$ were classified as showing reliable change beyond measurement error.

\subsection*{Statistical Analysis}
\textbf{Primary effects.} We estimated causal effects using linear mixed-effects models with random intercepts for participants to account for repeated measurements:
\[
GIH_{it} = \beta_0 + \beta_1 \text{Condition}_i + \beta_2 \text{Time}_t + \beta_3 (\text{Condition}_i \times \text{Time}_t) + u_i + \epsilon_{it},
\]
where $GIH_{it}$ denotes the GIH score for participant $i$ at time $t$, $\text{Condition}_i$ indicates treatment assignment, $\text{Time}_t$ represents measurement occasion, $u_i$ is a participant-specific random intercept, and $\epsilon_{it}$ is the residual error term. The interaction coefficient $\beta_3$ captures the treatment effect over time.

\textbf{Difference-in-differences (DiD).} To quantify treatment effects at each post-baseline time point, we computed difference-in-differences estimates:
\[
\Delta\Delta =
(\bar{Y}_{T,t} - \bar{Y}_{T,\text{pre}})
-
(\bar{Y}_{C,t} - \bar{Y}_{C,\text{pre}}),
\]
where $\bar{Y}_{T,t}$ and $\bar{Y}_{C,t}$ denote mean GIH scores in the treatment and control groups at time $t$, respectively. 

\textbf{Moderation analyses.} We examined whether treatment effects varied by political affiliation and openness to experience, using ordinary least squares (OLS) regression models to predict individual change scores. Models included treatment condition, the moderator, their interaction, and baseline GIH as a covariate. Heteroscedasticity-robust (HC3) standard errors were used.

All statistical tests were two-sided.

\backmatter
\bmhead{Acknowledgments}
A faculty startup fund from the University of South Florida supported this work.

\bmhead{Conflicts of Interest} The authors declare no conflict of interest.

\bmhead{Ethics Approval} This research was approved by the IRB at the University of South Florida.

\bmhead{Data, Materials, and Software Availability} The data and code used in this paper are publicly available at \url{https://github.com/cssai-research/bloom-socratic-IH}.

\bmhead{Author Contributions} MRM designed the study, collected and analyzed the data, interpreted the results, and authored the manuscript. RAB designed the study, analyzed the data, interpreted the results, authored the manuscript, and supervised the research.

\clearpage

\begin{appendices}
\setcounter{figure}{0}
\setcounter{table}{0}
\renewcommand{\thefigure}{S\arabic{figure}}
\renewcommand{\thetable}{S\arabic{table}}

\makeatletter
\renewcommand{\theHfigure}{S\arabic{figure}}
\renewcommand{\theHtable}{S\arabic{table}}
\makeatother

\renewcommand{\thesection}{A.\arabic{section}}
\renewcommand{\thesubsection}{\thesection.\arabic{subsection}}
\renewcommand{\thesubsubsection}{\thesubsection.\arabic{subsubsection}}

\makeatletter
\renewcommand{\theHsection}{A.\arabic{section}}
\renewcommand{\theHsubsection}{\theHsection.\arabic{subsection}}
\renewcommand{\theHsubsubsection}{\theHsubsection.\arabic{subsubsection}}
\makeatother

\section*{Supplementary Materials for \textit{General Intellectual Humility Is Malleable Through AI-Mediated Reflective Dialogue}}\label{SM}

\setcounter{section}{0}
\renewcommand{\appendixname}{}

\section{LLM-based Intervention System Design}
\subsection{LLM Prompt for the Treatment Condition}
The intervention was implemented as a three-stage structured conversation, where each stage lasted approximately 4 minutes. The stage progression and timing were controlled externally by the user interface (UI), while the backend server dynamically selected stage-specific system prompts.

To reflect this architecture, the system prompt was decomposed into a shared base prompt and stage-specific augmentations. The stage was passed algorithmically from the UI to the server as \texttt{\{\{stage\}\}}, which determined which prompt variant was used.

\subsubsection*{Base Prompt (shared across all stages):}
\begin{quote}
``You are an AI facilitator guiding a human participant through a structured conversation to promote intellectual humility (IH). Your goal is to help the participant reflect on and learn three facets of IH:
\begin{enumerate}
    \item Awareness of limits of one’s own knowledge
    \item Openness to opposing viewpoints
    \item Willingness to revise beliefs based on strong evidence
\end{enumerate}

Adopt a conversational style that is friendly, accepting, and trustworthy, since people show greater intellectual humility when they feel understood and supported. \\

\{Stage-specific prompt augmentation selected via \texttt{stage}\} \\

Turn-by-turn instructions:
\begin{enumerate}
    \item Keep responses concise (approximately 20--40 words)
    \item Provide a brief IH-based response grounded in the participant’s previous answer
    \item Ask exactly one crisp, open-ended Socratic question that builds on the participant’s reflection
\end{enumerate}

Additional guidelines:
\begin{enumerate}
    \item Cover all three IH concepts (awareness, openness, willingness)
    \item Do not answer unrelated questions or let the conversation drift off-topic
    \item Use natural, conversational, and easy-to-understand language"
\end{enumerate}
\end{quote}

\subsubsection*{Stage-Specific Prompt Augmentation:}
\begin{itemize}
    \item \textbf{\{\{stage = stage1\}\}} (Remember \& Understand):
    \begin{quote}
    ``Specific goals of this stage: Participants will recall the key concepts of intellectual humility (awareness of one’s own knowledge limits, openness to opposing viewpoints, and willingness to revise beliefs based on evidence). They will also explain intellectual humility in their own words to demonstrate understanding.''
    \end{quote}

    \item \textbf{\{\{stage = stage2\}\}} (Apply \& Analyze):
    \begin{quote}
    ``Specific goals of this stage: Participants will apply intellectual humility in given scenarios based on key concepts of IH (awareness, openness, willingness). They will then analyze their own responses, making connections between different IH elements (e.g., confidence vs humility, facts vs opinions, personal bias vs objective reasoning).''
    \end{quote}

    \item \textbf{\{\{stage = stage3\}\}} (Evaluate \& Create):
    \begin{quote}
    ``Specific goals of this stage: Participants will create new scenarios or share original examples where intellectual humility is beneficial, or where its absence causes problems. Participants will justify their scenarios or examples and reasoning based on IH components (awareness, openness, willingness).'' \newpage
    \end{quote}
\end{itemize}

\subsubsection*{UI-Controlled Stage Progression:}
Each stage lasted \texttt{\{\{STAGE\_DURATION\}\}} seconds and was enforced by a client-side timer. After the timer expired, participants could either proceed to the next stage or continue the conversation voluntarily.
\begin{itemize}
    \item \texttt{\{\{STAGE\_DURATION\}\}} = 240 seconds (4 minutes)
    \item Total structured duration = 12 minutes
    \item Stage transitions were handled in the UI and communicated to the backend via \texttt{\{\{stage\}\}}
\end{itemize}

\subsubsection*{Dynamic Prompt Selection:}
At each interaction step, the backend received the full conversation history along with a stage identifier \texttt{\{\{stage\}\}} from the client. The server then dynamically selected the appropriate stage-specific system prompt:
\begin{quote}
\texttt{system\_prompt = getSystemPromptForStage(\{\{stage\}\})}
\end{quote}

The final input to the language model was constructed by prepending this system prompt to the conversation history:
\begin{quote}
\texttt{full\_messages = [system\_prompt, messages]}
\end{quote}

This design ensured that the model behavior was conditioned explicitly on the current stage of the intervention, stage transitions were controlled externally by the UI and passed to the backend via \texttt{\{\{stage\}\}}, and the language model itself remained stateless with respect to timing and stage progression.

\subsection{LLM Prompt for the Control Condition}
Participants in the control condition engaged in conversations guided by one of three topic-specific system prompts \cite{costello2024durably}. These prompts were designed to facilitate discussion or debate without explicitly targeting General Intellectual Humility.

Similar to the treatment condition, a stage-based prompting framework was used, where the current stage (\texttt{\{\{stage\}\}}) was passed from the client to the backend, and the corresponding system prompt was dynamically selected at each turn.

\paragraph*{Healthcare System Discussion}
\begin{quote}
``Engage with users about their experience with the American medical system. Your objective is to facilitate a discussion where the user can express and elaborate on their experiences and beliefs. Use simple language that an average person will be able to understand. Avoid discussing or leading the conversation toward conspiracy theories, politics, religion, or any potentially sensitive subjects. Use open-ended questions to encourage users to share their thoughts and experiences."
\end{quote}

\paragraph*{Pets: Cats vs.\ Dogs}
\begin{quote}
``Your objective is to debate with users about whether cats or dogs are better. This is an exercise in disagreement and debate. You should probe the key points of the user's argument, and perspective, and find points of argument. Use simple language that an average person will be able to understand. Avoid discussing or leading the conversation toward conspiracy theories, politics, religion, or any potentially sensitive subjects."
\end{quote}

\paragraph*{Firefighters}
\begin{quote}
``Engage with users about their experience with firefighters. Your objective is to facilitate a discussion where the user can express and elaborate on their experiences and beliefs. Use simple language that an average person will be able to understand. Avoid discussing or leading the conversation toward conspiracy theories, politics, religion, or any potentially sensitive subjects. Use open-ended questions to encourage users to share their thoughts and experiences."
\end{quote}

\section{Survey Instruments}
\subsection{Demographic Questionnaire}
Participants completed the following demographic questionnaire:

\begin{enumerate}
    \item \textbf{Age range}
    \begin{itemize}
        \item 18--24 years old
        \item 25--34 years old
        \item 35--44 years old
        \item 45--54 years old
        \item 55--64 years old
        \item 65+ years old
    \end{itemize}

    \item \textbf{Race (select all that apply)}
    \begin{itemize}
        \item White
        \item Black or African American
        \item American Indian or Alaska Native
        \item Asian
        \item Native Hawaiian or Other Pacific Islander
        \item Some other race
        \item I prefer not to answer
    \end{itemize}

    \item \textbf{Are you of Hispanic or Latino origin?}
    \begin{itemize}
        \item Yes
        \item No
        \item I prefer not to answer
    \end{itemize}

    \item \textbf{Gender}
    \begin{itemize}
        \item Male
        \item Female
        \item Non-binary
        \item I prefer not to answer
    \end{itemize}

    \item \textbf{Highest degree of education}
    \begin{itemize}
        \item Less than a high school degree
        \item High school degree or equivalent (e.g., GED)
        \item Some college but no degree
        \item Associate’s degree (occupational or academic)
        \item Bachelor’s degree
        \item Master’s degree
        \item Professional degree
        \item Doctoral degree
        \item Some other degree
        \item I prefer not to answer
    \end{itemize}

    \item \textbf{Occupation}
    \begin{itemize}
        \item Open-ended text response
    \end{itemize}

    \item \textbf{Party affiliation}
    \begin{itemize}
        \item Democrat
        \item Republican
        \item Independent
        \item Some other party
        \item Prefer not to say
    \end{itemize}
\end{enumerate}

\subsection{General Intellectual Humility Scale}

The items are rated on a 7-point Likert scale with
\emph{1 = Strongly Disagree},
\emph{2 = Disagree},
\emph{3 = Slightly Disagree},
\emph{4 = Neither Agree nor Disagree},
\emph{5 = Slightly Agree},
\emph{6 = Agree},
\emph{7 = Strongly Agree}.

\begin{enumerate}
    \item I question my own opinions, positions, and viewpoints because they could be wrong.
    \item I reconsider my opinions when presented with new evidence.
    \item I recognize the value in opinions that are different from my own.
    \item I accept that my beliefs and attitudes may be wrong.
    \item In the face of conflicting evidence, I am open to changing my opinions.
    \item I am open to finding out new information that differs from what I already think is true.
\end{enumerate}

\subsection{Big Five Inventory--2 Short Form (BFI-2-S)}
The Big Five Inventory--2 Short Form (BFI-2-S)~\cite{soto2017short,soto2017next} was used to assess personality traits across the five broad dimensions of personality. Participants were instructed to indicate the extent to which they agreed with each statement using a 5-point Likert scale, with
\emph{1 = Disagree strongly},
\emph{2 = Disagree a little},
\emph{3 = Neutral; no opinion},
\emph{4 = Agree a little}, and
\emph{5 = Agree strongly}.

\noindent Participants responded to the following prompt: \emph{``I am someone who\ldots''}

\begin{enumerate}
    \item Tends to be quiet.
    \item Is compassionate, has a soft heart.
    \item Tends to be disorganized.
    \item Worries a lot.
    \item Is fascinated by art, music, or literature.
    \item Is dominant, acts as a leader.
    \item Is sometimes rude to others.
    \item Has difficulty getting started on tasks.
    \item Tends to feel depressed, blue.
    \item Has little interest in abstract ideas.
    \item Is full of energy.
    \item Assumes the best about people.
    \item Is reliable, can always be counted on.
    \item Is emotionally stable, not easily upset.
    \item Is original, comes up with new ideas.
    \item Is outgoing, sociable.
    \item Can be cold and uncaring.
    \item Keeps things neat and tidy.
    \item Is relaxed, handles stress well.
    \item Has few artistic interests.
    \item Prefers to have others take charge.
    \item Is respectful, treats others with respect.
    \item Is persistent, works until the task is finished.
    \item Feels secure, comfortable with self.
    \item Is complex, a deep thinker.
    \item Is less active than other people.
    \item Tends to find fault with others.
    \item Can be somewhat careless.
    \item Is temperamental, gets emotional easily.
    \item Has little creativity.
\end{enumerate}

\section{Supplementary Tables}
\begin{table}[ht]
\centering
\caption{Demographic characteristics by condition (Treatment: $N = 200$; Control: $N = 200$)}
\label{tab:s_demographics}
\begin{tabular*}{\textwidth}{@{\extracolsep{\fill}}llcc}
\toprule
Category & Group & Control ($N$) & Treatment ($N$) \\
\midrule

\multicolumn{4}{l}{\textit{Age}} \\
& 18--24 & 13 & 10 \\
& 25--34 & 51 & 65 \\
& 35--44 & 56 & 53 \\
& 45--54 & 42 & 50 \\
& 55--64 & 28 & 17 \\
& 65+ & 10 & 5 \\

\addlinespace
\multicolumn{4}{l}{\textit{Gender}} \\
& Female & 106 & 99 \\
& Male & 91 & 97 \\
& Non-binary & 2 & 3 \\
& Prefer not to answer & 1 & 1 \\

\addlinespace
\multicolumn{4}{l}{\textit{Education}} \\
& Bachelor's degree & 78 & 73 \\
& Master's degree & 34 & 36 \\
& Some college & 30 & 32 \\
& Associate degree & 24 & 22 \\
& High school graduate & 22 & 30 \\
& Doctorate degree & 8 & 4 \\
& Professional degree & 2 & 2 \\
& Less than high school & 2 & 1 \\

\addlinespace
\multicolumn{4}{l}{\textit{Political affiliation}} \\
& Democrat & 85 & 99 \\
& Republican & 60 & 43 \\
& Independent & 48 & 51 \\
& Other & 3 & 4 \\
& Prefer not to answer & 4 & 3 \\

\addlinespace
\multicolumn{4}{l}{\textit{Race}} \\
& White & 142 & 134 \\
& Black or African American & 27 & 35 \\
& Asian & 15 & 13 \\
& American Indian or Alaska Native & 1 & 2 \\
& Other & 6 & 5 \\
& Multiracial / other combinations & 9 & 11 \\

\addlinespace
\multicolumn{4}{l}{\textit{Hispanic origin}} \\
& No & 183 & 182 \\
& Yes & 16 & 17 \\
& Prefer not to answer & 1 & 1 \\

\bottomrule
\end{tabular*}
\end{table}

\begin{table}[ht]
\centering
\caption{Linear mixed-effects model predicting general intellectual humility (pre vs. post)}
\label{tab:mixed_model_GIH_prepost}
\begin{tabular*}{\textwidth}{@{\extracolsep{\fill}}lccccc}
\toprule
Predictor & Estimate ($\beta$) & SE & $z$ & 95\% CI & $p$ \\
\midrule
Intercept & 5.902 & 0.052 & 113.092 & $[5.799,\ 6.004]$ & $<.001$ \\
Time (Post vs. Pre) & 0.073 & 0.028 & 2.59 & $[0.018,\ 0.129]$ & .010 \\
Treatment (vs. Control) & $-0.026$ & 0.074 & $-0.35$ & $[-0.170,\ 0.119]$ & .726 \\
Treatment $\times$ Time & 0.187 & 0.040 & 4.67 & $[0.108,\ 0.265]$ & $<.001$ \\
\bottomrule
\end{tabular*}
\end{table}

\begin{table}[ht]
\centering
\caption{Linear mixed-effects model predicting general intellectual humility across three timepoints}
\label{tab:mixed_model_GIH_followup}
\begin{tabular*}{\textwidth}{@{\extracolsep{\fill}}lccccc}
\toprule
Predictor & Estimate ($\beta$) & SE & $z$ & 95\% CI & $P$ \\
\midrule
Intercept & 5.902 & 0.052 & 113.863 & $[5.800,\ 6.003]$ & $<.001$ \\
Time (Post vs. Pre) & 0.073 & 0.031 & 2.345 & $[0.012,\ 0.135]$ & .019 \\
Time (Follow-up vs. Pre) & $-0.045$ & 0.033 & $-1.355$ & $[-0.110,\ 0.020]$ & .175 \\
Treatment (vs. Control) & $-0.026$ & 0.073 & $-0.352$ & $[-0.170,\ 0.118]$ & .725 \\
Treatment $\times$ Post & 0.187 & 0.044 & 4.22 & $[0.100,\ 0.273]$ & $<.001$ \\
Treatment $\times$ Follow-up & 0.187 & 0.047 & 4.01 & $[0.096,\ 0.279]$ & $<.001$ \\
\bottomrule
\end{tabular*}
\end{table}

\begin{table}[ht]
\centering
\caption{Heteroscedasticity-robust OLS regression predicting change in general intellectual humility ($\Delta$GIH) with political affiliation moderation}
\label{tab:party_moderation}
\begin{tabular*}{\textwidth}{@{\extracolsep{\fill}}lccccc}
\toprule
Predictor & Estimate ($\beta$) & SE & $z$ & 95\% CI & $P$ \\
\midrule
Intercept & 1.391 & 0.212 & 6.55 & $[0.975,\ 1.807]$ & $<.001$ \\
Treatment (vs. Control) & 0.157 & 0.052 & 3.00 & $[0.054,\ 0.259]$ & .003 \\
Independent (vs. Democrat) & $-0.020$ & 0.053 & $-0.37$ & $[-0.124,\ 0.085]$ & .710 \\
Other/Unknown (vs. Democrat) & 0.052 & 0.138 & 0.38 & $[-0.218,\ 0.322]$ & .706 \\
Republican (vs. Democrat) & $-0.158$ & 0.067 & $-2.37$ & $[-0.288,\ -0.027]$ & .018 \\
Treatment $\times$ Independent & $-0.022$ & 0.086 & $-0.26$ & $[-0.191,\ 0.146]$ & .796 \\
Treatment $\times$ Other/Unknown & 0.066 & 0.185 & 0.36 & $[-0.296,\ 0.428]$ & .720 \\
Treatment $\times$ Republican & 0.068 & 0.104 & 0.65 & $[-0.136,\ 0.272]$ & .516 \\
Baseline GIH (pre) & $-0.215$ & 0.034 & $-6.36$ & $[-0.281,\ -0.149]$ & $<.001$ \\
\bottomrule
\end{tabular*}
\end{table}

\begin{table}[ht]
\centering
\caption{Heteroscedasticity-robust OLS models testing moderation of treatment effects by Big Five personality traits}
\label{tab:big5_moderation}
\begin{tabular*}{\textwidth}{@{\extracolsep{\fill}}lcccccc}
\toprule
 & \multicolumn{3}{c}{Trait Main Effect} & \multicolumn{3}{c}{Treatment $\times$ Trait} \\
\cmidrule(lr){2-4} \cmidrule(lr){5-7}
Trait & $\beta$ & 95\% CI & $P$ & $\beta$ & 95\% CI & $P$ \\
\midrule
Openness & 0.093 & $[0.021,\ 0.165]$ & .012 & $-0.072$ & $[-0.187,\ 0.043]$ & .217 \\
Agreeableness & 0.031 & $[-0.023,\ 0.084]$ & .266 & $-0.003$ & $[-0.087,\ 0.080]$ & .940 \\
Conscientiousness & 0.021 & $[-0.030,\ 0.071]$ & .417 & 0.057 & $[-0.014,\ 0.127]$ & .114 \\
Extraversion & $-0.001$ & $[-0.046,\ 0.045]$ & .977 & 0.031 & $[-0.035,\ 0.096]$ & .358 \\
Neuroticism & 0.020 & $[-0.021,\ 0.062]$ & .330 & $-0.047$ & $[-0.109,\ 0.015]$ & .137 \\
\bottomrule
\end{tabular*}
\end{table}

\begin{table}[ht]
\centering
\caption{Correlations between baseline general intellectual humility (GIH) and Big Five personality traits}
\label{tab:big5_gih_correlations}
\begin{tabular*}{\textwidth}{@{\extracolsep{\fill}}lcccccccc}
\toprule
& \multicolumn{3}{c}{Pearson} & \multicolumn{2}{c}{Spearman} & \multicolumn{2}{c}{Kendall} \\
\cmidrule(lr){2-4} \cmidrule(lr){5-6} \cmidrule(lr){7-8}
Trait & $r$ & 95\% CI & $P$ & $\rho$ & $P$ & $\tau$ & $P$ \\
\midrule
Openness & 0.230 & $[0.135,\ 0.321]$ & $<.001$ & 0.283 & $<.001$ & 0.208 & $<.001$ \\
Agreeableness & 0.139 & $[0.041,\ 0.234]$ & .005 & 0.205 & $<.001$ & 0.147 & $<.001$ \\
Conscientiousness & 0.070 & $[-0.029,\ 0.166]$ & .165 & 0.125 & .013 & 0.089 & .013 \\
Extraversion & 0.078 & $[-0.020,\ 0.175]$ & .118 & 0.095 & .057 & 0.065 & .066 \\
Neuroticism & $-0.024$ & $[-0.122,\ 0.074]$ & .630 & $-0.092$ & .067 & $-0.062$ & .082 \\
\bottomrule
\end{tabular*}
\end{table}

\begin{table}[ht]
\centering
\caption{Heteroscedasticity-robust OLS regression predicting change in general intellectual humility ($\Delta$GIH) with covariate adjustment}
\label{tab:covariate_model_big5}
\begin{tabular*}{\textwidth}{@{\extracolsep{\fill}}lccccc}
\toprule
Predictor & Estimate ($\beta$) & SE & $z$ & 95\% CI & $P$ \\
\midrule
Intercept & 0.607 & 0.290 & 2.09 & $[0.038,\ 1.176]$ & .037 \\
Treatment (vs. Control) & 0.194 & 0.037 & 5.22 & $[0.121,\ 0.267]$ & $<.001$ \\
Baseline GIH (pre) & $-0.211$ & 0.037 & $-5.65$ & $[-0.285,\ -0.138]$ & $<.001$ \\
Duration (seconds) & $2.01 \times 10^{-5}$ & $4.56 \times 10^{-5}$ & 0.44 & $[-6.93\times10^{-5},\ 0.000]$ & .659 \\
User turns & 0.005 & 0.003 & 1.79 & $[-0.001,\ 0.011]$ & .074 \\
Openness & 0.057 & 0.031 & 1.82 & $[-0.005,\ 0.119]$ & .069 \\
Agreeableness & 0.008 & 0.026 & 0.30 & $[-0.044,\ 0.059]$ & .768 \\
Conscientiousness & 0.078 & 0.027 & 2.90 & $[0.025,\ 0.131]$ & .004 \\
Extraversion & $-0.016$ & 0.023 & $-0.69$ & $[-0.062,\ 0.030]$ & .491 \\
Neuroticism & 0.030 & 0.023 & 1.29 & $[-0.016,\ 0.076]$ & .197 \\
\bottomrule
\end{tabular*}
\end{table}

\clearpage
\end{appendices}

\bibliography{references}

@article{krumrei2016development,
  title={The development and validation of the comprehensive intellectual humility scale},
  author={Krumrei-Mancuso, Elizabeth J and Rouse, Steven V},
  journal={Journal of Personality Assessment},
  volume={98},
  number={2},
  pages={209--221},
  year={2016},
  publisher={Taylor \& Francis}
}

@article{soto2017short,
  title={Short and extra-short forms of the {Big Five Inventory}--2: The {BFI-2-S} and {BFI-2-XS}},
  author={Soto, Christopher J and John, Oliver P},
  journal={Journal of Research in Personality},
  volume={68},
  pages={69--81},
  year={2017},
  publisher={Elsevier}
}

@article{soto2017next,
  title={The next {Big Five Inventory (BFI-2)}: Developing and assessing a hierarchical model with 15 facets to enhance bandwidth, fidelity, and predictive power},
  author={Soto, Christopher J and John, Oliver P},
  journal={Journal of Personality and Social Psychology},
  volume={113},
  number={1},
  pages={117},
  year={2017},
  publisher={American Psychological Association}
}

@article{fleeson2004moving,
  title={Moving personality beyond the person-situation debate: The challenge and the opportunity of within-person variability},
  author={Fleeson, William},
  journal={Current Directions in Psychological Science},
  volume={13},
  number={2},
  pages={83--87},
  year={2004},
  publisher={SAGE Publications Sage CA: Los Angeles, CA}
}

@article{roberts2005evaluating,
  title={Evaluating five factor theory and social investment perspectives on personality trait development},
  author={Roberts, Brent W and Wood, Dustin and Smith, Jennifer L},
  journal={Journal of Research in Personality},
  volume={39},
  number={1},
  pages={166--184},
  year={2005},
  publisher={Elsevier}
}

@article{roberts2008personality,
  title={Personality trait change in adulthood},
  author={Roberts, Brent W and Mroczek, Daniel},
  journal={Current Directions in Psychological Science},
  volume={17},
  number={1},
  pages={31--35},
  year={2008},
  publisher={SAGE Publications Sage CA: Los Angeles, CA}
}

@article{porter2018intellectual,
  title={Intellectual humility and openness to the opposing view},
  author={Porter, Tenelle and Schumann, Karina},
  journal={Self and Identity},
  volume={17},
  number={2},
  pages={139--162},
  year={2018},
  publisher={Taylor \& Francis}
}

@article{porter2022predictors,
  title={Predictors and consequences of intellectual humility},
  author={Porter, Tenelle and Elnakouri, Abdo and Meyers, Ethan A and Shibayama, Takuya and Jayawickreme, Eranda and Grossmann, Igor},
  journal={Nature Reviews Psychology},
  volume={1},
  number={9},
  pages={524--536},
  year={2022},
  publisher={Nature Publishing Group US New York}
}

@article{kross2012boosting,
  title={Boosting wisdom: Distance from the self enhances wise reasoning, attitudes, and behavior},
  author={Kross, Ethan and Grossmann, Igor},
  journal={Journal of Experimental Psychology: General},
  volume={141},
  number={1},
  pages={43},
  year={2012},
  publisher={American Psychological Association}
}

@article{hanel2023using,
  title={Using self-affirmation to increase intellectual humility in debate},
  author={Hanel, Paul HP and Roy, Deborah and Taylor, Samuel and Franjieh, Michael and Heffer, Chris and Tanesini, Alessandra and Maio, Gregory R},
  journal={Royal Society Open Science},
  volume={10},
  number={2},
  year={2023},
  pages   = {220958},
  doi     = {10.1098/rsos.220958},
  publisher={The Royal Society}
}

@article{meagher2019intellectually,
  title={An intellectually humbling experience: Changes in interpersonal perception and cultural reasoning across a five-week course},
  author={Meagher, Benjamin R and Gunn, Hanna and Sheff, Nathan and Van Tongeren, Daryl R},
  journal={Journal of Psychology and Theology},
  volume={47},
  number={3},
  pages={217--229},
  year={2019},
  publisher={Sage Publications Sage UK: London, England}
}

@article{anderson2021development,
  title={The development of intellectual humility as an impact of a week-long philosophy summer camp for teens and tweens: Preliminary results},
  author={Anderson, David J and Holte, Patricia N and Maffly-Kipp, Joseph and Conway, Daniel and Katz, Claire Elise and Schlegel, Rebecca J},
  journal={Precollege Philosophy and Public Practice},
  volume={3},
  pages={41--65},
  year={2021}
}

@article{thorson2025increases,
  title={Increases in intellectual humility from guided conversations are greater when people perceive affiliation with conversation partners},
  author={Thorson, Katherine R and Beck, Lindsey A and Ketay, Sarah and Welker, Keith M},
  journal={Social Psychological and Personality Science},
  volume={16},
  number={3},
  pages={313--323},
  year={2025},
  publisher={SAGE Publications Sage CA: Los Angeles, CA}
}

@article{king1994guiding,
  title={Guiding knowledge construction in the classroom: Effects of teaching children how to question and how to explain},
  author={King, Alison},
  journal={American Educational Research Journal},
  volume={31},
  number={2},
  pages={338--368},
  year={1994},
  publisher={Sage Publications}
}

@book{vlastos1994socratic,
  author    = {Vlastos, Gregory},
  editor    = {Burnyeat, Myles},
  title     = {Socratic Studies},
  year      = {1994},
  publisher = {Cambridge University Press},
  address   = {Cambridge},
  isbn      = {9780521447355},
}

@book{anderson2001taxonomy,
  editor    = {Anderson, Lorin W. and Krathwohl, David R.},
  title     = {A Taxonomy for Learning, Teaching, and Assessing: 
               A Revision of {Bloom}'s Taxonomy of Educational Objectives},
  year      = {2001},
  publisher = {Longman},
  address   = {New York},
}

@article{paul2007critical,
  title={Critical thinking: The art of {Socratic} questioning},
  author={Paul, Richard and Elder, Linda},
  journal={Journal of Developmental Education},
  volume={31},
  number={1},
  pages={36},
  year={2007},
  publisher={Appalachian State University d/b/a}
}

@book{bloom1956taxonomy,
  title={Taxonomy of Educational Objectives: The Classification of Educational Goals. {Handbook} I: Cognitive Domain},
  author={Bloom, Benjamin S. and Engelhart, Max D. and Furst, Edward J. and Hill, Walker H. and Krathwohl, David R.},
  year={1956},
  publisher={Longmans, Green},
  address={New York}
}

@article{krumrei2025toward,
  title={Toward an understanding of collective intellectual humility},
  author={Krumrei-Mancuso, Elizabeth J and P{\"a}rnamets, Philip and Bland, Steven and Astola, Mandi and Cichocka, Aleksandra and de Ridder, Jeroen and Mercier, Hugo and Meyer, Marco and O’connor, Cailin and Porter, Tenelle and others},
  journal={Trends in Cognitive Sciences},
  volume={29},
  number={1},
  pages={15--27},
  year={2025},
  publisher={Elsevier}
}

@article{porter2020intellectual,
  title={Intellectual humility predicts mastery behaviors when learning},
  author={Porter, Tenelle and Schumann, Karina and Selmeczy, Diana and Trzesniewski, Kali},
  journal={Learning and Individual Differences},
  volume={80},
  pages={101888},
  year={2020},
  doi     = {10.1016/j.lindif.2020.101888},
  publisher={Elsevier}
}

@article{stieger2018peach,
  title={{PEACH}, a smartphone-and conversational agent-based coaching intervention for intentional personality change: study protocol of a randomized, wait-list controlled trial},
  author={Stieger, Mirjam and Ni{\ss}en, Marcia and R{\"u}egger, Dominik and Kowatsch, Tobias and Fl{\"u}ckiger, Christoph and Allemand, Mathias},
  journal={BMC Psychology},
  volume={6},
  number={1},
  pages={43},
  year={2018},
  publisher={Springer}
}

@inproceedings{kim2025conversational,
  author    = {Kim, Antino and Dennis, Alan R. and Smith, Evelyn O.},
  title     = {Conversational Agents Powered by Generative Artificial Intelligence Improve Understanding for Those Worried About Controversial Topics},
  booktitle = {Proceedings of the 58th Hawaii International Conference on System Sciences (HICSS)},
  year      = {2025},
  pages     = {2698--2704},
  publisher = {Hawaii International Conference on System Sciences},
  address   = {Waikoloa, Hawaii}
}

@article{costello2024durably,
  title={Durably reducing conspiracy beliefs through dialogues with {AI}},
  author={Costello, Thomas H and Pennycook, Gordon and Rand, David G},
  journal={Science},
  volume={385},
  number={6714},
  pages={eadq1814},
  year={2024},
  publisher={American Association for the Advancement of Science}
}

@article{argyle2023leveraging,
  title={Leveraging {AI} for democratic discourse: Chat interventions can improve online political conversations at scale},
  author={Argyle, Lisa P and Bail, Christopher A and Busby, Ethan C and Gubler, Joshua R and Howe, Thomas and Rytting, Christopher and Sorensen, Taylor and Wingate, David},
  journal={Proceedings of the National Academy of Sciences},
  volume={120},
  number={41},
  pages={e2311627120},
  year={2023},
  publisher={National Academy of Sciences}
}

@article{smith2023you,
  title={You know you're right: How intellectual humility decreases political hostility},
  author={Smith, Glen},
  journal={Political Psychology},
  volume={44},
  number={6},
  pages={1319--1335},
  year={2023},
  publisher={Wiley Online Library}
}

@article{krumrei2020intellectual,
  title={Intellectual humility in the sociopolitical domain},
  author={Krumrei-Mancuso, Elizabeth J and Newman, Brian},
  journal={Self and Identity},
  volume={19},
  number={8},
  pages={989--1016},
  year={2020},
  publisher={Taylor \& Francis}
}

@article{grossmann2014exploring,
  title={Exploring {Solomon}’s paradox: Self-distancing eliminates the self-other asymmetry in wise reasoning about close relationships in younger and older adults},
  author={Grossmann, Igor and Kross, Ethan},
  journal={Psychological Science},
  volume={25},
  number={8},
  pages={1571--1580},
  year={2014},
  publisher={Sage Publications Sage CA: Los Angeles, CA}
}

@article{knochelmann2024effects,
  author  = {Kn{\"o}chelmann, Larissa and Cohrs, J. Christopher},
  title   = {Effects of intellectual humility in the context of affective polarization: Approaching and avoiding others in controversial political discussions},
  journal = {Journal of Personality and Social Psychology},
  year    = {2025},
  volume  = {129},
  number  = {1},
  pages   = {91--117},
  doi     = {10.1037/pspi0000462},
}

@article{mellers2019forecasting,
  title={Forecasting tournaments, epistemic humility and attitude depolarization},
  author={Mellers, Barbara and Tetlock, Philip and Arkes, Hal R},
  journal={Cognition},
  volume={188},
  pages={19--26},
  year={2019},
  doi     = {10.1016/j.cognition.2018.10.021},
  publisher={Elsevier}
}

@article{stieger2021changing,
  title={Changing personality traits with the help of a digital personality change intervention},
  author={Stieger, Mirjam and Fl{\"u}ckiger, Christoph and R{\"u}egger, Dominik and Kowatsch, Tobias and Roberts, Brent W and Allemand, Mathias},
  journal={Proceedings of the National Academy of Sciences},
  volume={118},
  number={8},
  pages={e2017548118},
  year={2021},
  publisher={National Academy of Sciences}
}

@article{spiegel2012open,
  title={Open-mindedness and intellectual humility},
  author={Spiegel, James S},
  journal={Theory and Research in Education},
  volume={10},
  number={1},
  pages={27--38},
  year={2012},
  publisher={SAGE Publications Sage UK: London, England}
}

@article{duckworth2015measurement,
  title={Measurement matters: Assessing personal qualities other than cognitive ability for educational purposes},
  author={Duckworth, Angela L and Yeager, David Scott},
  journal={Educational Researcher},
  volume={44},
  number={4},
  pages={237--251},
  year={2015},
  publisher={Sage Publications Sage CA: Los Angeles, CA}
}

@article{grossmann2021training,
  title={Training for wisdom: The distanced-self-reflection diary method},
  author={Grossmann, Igor and Dorfman, Anna and Oakes, Harrison and Santos, Henri C and Vohs, Kathleen D and Scholer, Abigail A},
  journal={Psychological Science},
  volume={32},
  number={3},
  pages={381--394},
  year={2021},
  publisher={Sage Publications Sage CA: Los Angeles, CA}
}

@article{walters2017known,
  title={Known unknowns: A critical determinant of confidence and calibration},
  author={Walters, Daniel J and Fernbach, Philip M and Fox, Craig R and Sloman, Steven A},
  journal={Management Science},
  volume={63},
  number={12},
  pages={4298--4307},
  year={2017},
  publisher={INFORMS}
}

@article{johnson2016reflecting,
  title={Reflecting on explanatory ability: A mechanism for detecting gaps in causal knowledge.},
  author={Johnson, Dan R and Murphy, Meredith P and Messer, Riley M},
  journal={Journal of Experimental Psychology: General},
  volume={145},
  number={5},
  pages={573},
  year={2016},
  publisher={American Psychological Association}
}

@article{brienza2017social,
  title={Social class and wise reasoning about interpersonal conflicts across regions, persons and situations},
  author={Brienza, Justin P and Grossmann, Igor},
  journal={Proceedings of the Royal Society B: Biological Sciences},
  volume={284},
  number={1869},
  pages   = {20171870},
  year={2017},
  doi     = {10.1098/rspb.2017.1870},
  publisher={The Royal Society}
}

@article{fernbach2013political,
  title={Political extremism is supported by an illusion of understanding},
  author={Fernbach, Philip M and Rogers, Todd and Fox, Craig R and Sloman, Steven A},
  journal={Psychological Science},
  volume={24},
  number={6},
  pages={939--946},
  year={2013},
  publisher={Sage Publications Sage CA: Los Angeles, CA}
}

@article{priest2017intellectual,
  title={Intellectual humility: An interpersonal theory},
  author={Priest, Maura},
  journal={Ergo, an Open Access Journal of Philosophy},
  volume={4},
  number  = {16},
  pages   = {463--480},
  doi     = {10.3998/ergo.12405314.0004.016},
  year={2017},
  publisher={Michigan Publishing, University of Michigan Library}
}

@article{meyers2020inducing,
  title={Inducing feelings of ignorance makes people more receptive to expert (economist) opinion},
  author={Meyers, Ethan A and Turpin, Martin H and Bia{\l}ek, Micha{\l} and Fugelsang, Jonathan A and Koehler, Derek J},
  journal={Judgment and Decision Making},
  volume={15},
  number={6},
  pages={909--925},
  year={2020},
  publisher={Cambridge University Press}
}

@article{grossmann2016wise,
  title={Wise reasoning in the face of everyday life challenges},
  author={Grossmann, Igor and Gerlach, Tanja M and Denissen, Jaap JA},
  journal={Social Psychological and Personality Science},
  volume={7},
  number={7},
  pages={611--622},
  year={2016},
  publisher={Sage Publications Sage CA: Los Angeles, CA}
}

@article{brienza2018wisdom,
  title={Wisdom, bias, and balance: Toward a process-sensitive measurement of wisdom-related cognition},
  author={Brienza, Justin P and Kung, Franki YH and Santos, Henri C and Bobocel, D Ramona and Grossmann, Igor},
  journal={Journal of Personality and Social Psychology},
  volume={115},
  number={6},
  pages={1093},
  year={2018},
  publisher={American Psychological Association}
}

@article{zachry2018situation,
  title={Situation-based contingencies underlying wisdom-content manifestations: Examining intellectual humility in daily life},
  author={Zachry, Corinne E and Phan, Le Vy and Blackie, Laura ER and Jayawickreme, Eranda},
  journal={The Journals of Gerontology: Series B},
  volume={73},
  number={8},
  pages={1404--1415},
  year={2018},
  publisher={Oxford University Press US}
}

@article{abrahams2019social,
  title={Social-emotional skill assessment in children and adolescents: Advances and challenges in personality, clinical, and educational contexts.},
  author={Abrahams, Loes and Pancorbo, Gina and Primi, Ricardo and Santos, Daniel and Kyllonen, Patrick and John, Oliver P and De Fruyt, Filip},
  journal={Psychological Assessment},
  volume={31},
  number={4},
  pages={460},
  year={2019},
  publisher={American Psychological Association}
}

@article{meagher2021intellectual,
  title={Intellectual humility in conversation: Distinct behavioral indicators of self and peer ratings},
  author={Meagher, Benjamin R and Leman, Joseph C and Heidenga, Caitlyn A and Ringquist, Michala R and Rowatt, Wade C},
  journal={The Journal of Positive Psychology},
  volume={16},
  number={3},
  pages={417--429},
  year={2021},
  publisher={Taylor \& Francis}
}

@article{danovitch2019intelligence,
  title={Intelligence and neurophysiological markers of error monitoring relate to children's intellectual humility},
  author={Danovitch, Judith H and Fisher, Megan and Schroder, Hans and Hambrick, David Z and Moser, Jason},
  journal={Child Development},
  volume={90},
  number={3},
  pages={924--939},
  year={2019},
  publisher={Oxford University Press}
}

@article{porter2022clarifying,
  title={Clarifying the content of intellectual humility: A systematic review and integrative framework},
  author={Porter, Tenelle and Baldwin, Chayce R and Warren, Michael T and Murray, Elise D and Cotton Bronk, Kendall and Forgeard, Marie JC and Snow, Nancy E and Jayawickreme, Eranda},
  journal={Journal of Personality Assessment},
  volume={104},
  number={5},
  pages={573--585},
  year={2022},
  publisher={Taylor \& Francis}
}

@article{hoekstra2021aspiring,
  title={Aspiring to greater intellectual humility in science},
  author={Hoekstra, Rink and Vazire, Simine},
  journal={Nature Human Behaviour},
  volume={5},
  number={12},
  pages={1602--1607},
  year={2021},
  publisher={Nature Publishing Group UK London}
}

@article{wong2021exploring,
  title={Exploring the relationship between intellectual humility and academic performance among post-secondary students: The mediating roles of learning motivation and receptivity to feedback},
  author={Wong, Ivy HM and Wong, Terry TY},
  journal={Learning and Individual Differences},
  volume={88},
  pages={102012},
  year={2021},
  doi     = {10.1016/j.lindif.2021.102012},
  publisher={Elsevier}
}

@article{bowes2021looking,
  title={Looking under the tinfoil hat: Clarifying the personological and psychopathological correlates of conspiracy beliefs},
  author={Bowes, Shauna M and Costello, Thomas H and Ma, Winkie and Lilienfeld, Scott O},
  journal={Journal of Personality},
  volume={89},
  number={3},
  pages={422--436},
  year={2021},
  publisher={Wiley Online Library}
}

@article{zmigrod2019psychological,
  title={The psychological roots of intellectual humility: The role of intelligence and cognitive flexibility},
  author={Zmigrod, Leor and Zmigrod, Sharon and Rentfrow, Peter Jason and Robbins, Trevor W},
  journal={Personality and Individual Differences},
  volume={141},
  pages={200--208},
  year={2019},
  doi     = {10.1016/j.paid.2018.12.007},
  publisher={Elsevier}
}

@article{krathwohl2002revision,
  title={A revision of {Bloom}'s taxonomy: An overview},
  author={Krathwohl, David R},
  journal={Theory into Practice},
  volume={41},
  number={4},
  pages={212--218},
  year={2002},
  publisher={Taylor \& Francis}
}

@article{scheffer2022belief,
  title={Belief traps: Tackling the inertia of harmful beliefs},
  author={Scheffer, Marten and Borsboom, Denny and Nieuwenhuis, Sander and Westley, Frances},
  journal={Proceedings of the National Academy of Sciences},
  volume={119},
  number={32},
  pages={e2203149119},
  year={2022},
  publisher={National Academy of Sciences}
}

@article{crawford2021asking,
  title={Asking people to explain complex policies does not increase political moderation: Three preregistered failures to closely replicate {Fernbach}, {Rogers}, {Fox}, and {Sloman}’s (2013) findings},
  author={Crawford, Jarret T and Ruscio, John},
  journal={Psychological Science},
  volume={32},
  number={4},
  pages={611--621},
  year={2021},
  publisher={Sage Publications Sage CA: Los Angeles, CA}
}

@inproceedings{rieger2024potential,
  title={From potential to practice: Intellectual humility during search on debated topics},
  author={Rieger, Alisa and Bredius, Frank and Theune, Mari{\"e}t and Pera, Maria Soledad},
  booktitle={Proceedings of the 2024 conference on human information interaction and retrieval},
  pages={130--141},
  year={2024}
}

@article{jackson2012can,
  title={Can an old dog learn (and want to experience) new tricks? {C}ognitive training increases openness to experience in older adults.},
  author={Jackson, Joshua J and Hill, Patrick L and Payne, Brennan R and Roberts, Brent W and Stine-Morrow, Elizabeth AL},
  journal={Psychology and Aging},
  volume={27},
  number={2},
  pages={286},
  year={2012},
  publisher={American Psychological Association}
}

@book{church2016intellectual,
  author    = {Ian M. Church and Peter L. Samuelson},
  title     = {Intellectual Humility: An Introduction to the Philosophy and Science},
  year      = {2016},
  publisher = {Bloomsbury Publishing},
  address   = {London},
}

@incollection{mccrae2009openness,
  author    = {McCrae, Robert R. and Sutin, Angelina R.},
  title     = {Openness to Experience},
  booktitle = {Handbook of Individual Differences in Social Behavior},
  editor    = {Leary, Mark R. and Hoyle, Rick H.},
  pages     = {257--273},
  year      = {2009},
  publisher = {Guilford Press},
  address   = {New York}
}

@article{leary2017cognitive,
  title={Cognitive and interpersonal features of intellectual humility},
  author={Leary, Mark R and Diebels, Kate J and Davisson, Erin K and Jongman-Sereno, Katrina P and Isherwood, Jennifer C and Raimi, Kaitlin T and Deffler, Samantha A and Hoyle, Rick H},
  journal={Personality and Social Psychology Bulletin},
  volume={43},
  number={6},
  pages={793--813},
  year={2017},
  publisher={Sage Publications Sage CA: Los Angeles, CA}
}

@article{proma2025exploring,
  title={Exploring the role of randomization on belief rigidity in online social networks},
  author={Proma, Adiba Mahbub and Pate, Neeley and Baten, Raiyan Abdul and Chen, Sifeng and Druckman, James N and Ghoshal, Gourab and Hoque, Ehsan},
  journal={IEEE Transactions on Affective Computing},
  year    = {2025},
  volume  = {13},
  number  = {1},
  pages   = {1171--1184},
  doi     = {10.1109/TAFFC.2025.3650069},
  publisher={IEEE}
}

@inproceedings{mahjabin2025wisdom,
  title={The Wisdom of Intellectually Humble Networks},
  author={Mahjabin, Mohammad Ratul and Baten, Raiyan Abdul},
  booktitle={Proceedings of the Annual Meeting of the Cognitive Science Society},
  volume={47},
  year={2025}
}

@article{hoyle2016holding,
  title={Holding specific views with humility: Conceptualization and measurement of specific intellectual humility},
  author={Hoyle, Rick H and Davisson, Erin K and Diebels, Kate J and Leary, Mark R},
  journal={Personality and Individual Differences},
  volume={97},
  pages={165--172},
  year={2016},
  doi     = {10.1016/j.paid.2016.03.043},
  publisher={Elsevier}
}

@article{gregg2014intellectual,
  title={Intellectual arrogance and intellectual humility: An evolutionary-epistemological account},
  author={Gregg, Aiden P and Mahadevan, Nikhila},
  journal={Journal of Psychology and Theology},
  volume={42},
  number={1},
  pages={7--18},
  year={2014},
  publisher={SAGE Publications Sage UK: London, England}
}

@article{bowes2022stepping,
  title={Stepping outside the echo chamber: Is intellectual humility associated with less political myside bias?},
  author={Bowes, Shauna M and Costello, Thomas H and Lee, Caroline and McElroy-Heltzel, Stacey and Davis, Don E and Lilienfeld, Scott O},
  journal={Personality and Social Psychology Bulletin},
  volume={48},
  number={1},
  pages={150--164},
  year={2022},
  publisher={Sage Publications Sage CA: Los Angeles, CA}
}

@incollection{church2017intellectual,
  author    = {Church, Ian M. and Barrett, Justin L.},
  title     = {Intellectual Humility},
  booktitle = {Handbook of Humility: Theory, Research, and Applications},
  editor    = {Worthington, Everett L., Jr. and Davis, Don E. and Hook, Joshua N.},
  pages     = {62--75},
  year      = {2017},
  publisher = {Routledge},
  address   = {New York}
}

@techreport{leary2018psychology,
  author      = {Leary, Mark R.},
  title       = {The Psychology of Intellectual Humility},
  institution = {John Templeton Foundation},
  year        = {2018},
  address     = {West Conshohocken, PA},
  url         = {https://www.templeton.org/wp-content/uploads/2018/11/Intellectual-Humility-Leary-FullLength-Final.pdf},
}

\end{document}